\documentclass{aastex}
\usepackage{spr-astr-addons}
\usepackage{url}

\RequirePackage{color}

\begin{document}

\title{String Black Holes as
Particle Accelerators to Arbitrarily High Energy}
\slugcomment{Not to appear in Non learned J., 45.}
\shorttitle{Short article title}
\shortauthors{Authors et al.}

\author{Parthapratim Pradhan}
\affil{Department of Physics, Vivekananda Satabarshiki Mahavidyalaya, Manikpara, West Midnapur, West Bengal~721513, India, Email: pppadhan77@gmail.com}


\begin{abstract}
We show that an  extremal Gibbons-Maeda-Garfinkle-Horowitz-Strominger black hole may act as a particle accelerator with arbitrarily high energy when two uncharged particles falling freely from rest to infinity on the near horizon. We show that the center of mass energy of collision is independent of the extreme  fine tuning of the angular momentum of the colliding particles. We further show that  the center of mass energy of collisions of particles  at the ISCO ($r_{ISCO}$) or at the photon orbit ($r_{ph}$) or at the marginally bound circular orbit ($r_{mb}$) i.e. at $r \equiv r_{ISCO}=r_{ph}=r_{mb}=2M$ could be arbitrarily large for the aforementioned space-time, which is quite different from the Schwarzschild and the Reissner-Nordstr{\o}m space-time. For non-extremal GMGHS space-time the CM energy is finite and depends upon the asymptotic value of the dilation field ($\phi_{0}$).
\end{abstract}

\keywords{BSW Effect, ISCO, Photon orbit.}

\section{Introduction}

An interesting effect predicted  by
\cite{bsw} (hereafter BSW) is that rotating black holes may act as particle accelerators
with arbitrarily high center-of-mass energy.
The process is likely: when two massive dark matter particles with
different angular momentum are falling into the black hole and collide near the horizon then the energy in the center of mass frame can become unlimited. Here it is assumed that the particles at infinity are at rest and the collision energy is produced mainly due to the gravitational acceleration. This mechanism is criticized by several authors. Particularly \cite{berti}, pointed out that there is an astrophysical limitations i.e. maximal spin, back reaction effect and gravitational radiation etc. on that center-of-mass(CM) energy due to the \cite{thorn}'s bound  i.e. $\frac{a}{M}=0.998$ ($M$ is the mass and $a$ is the spin of the black hole). Also \cite{jacob} found that CM energy in the near extremal situation for rotating Kerr black hole is $\frac{E_{cm}}{2m_{0}}\sim \frac{2.41}{(1-a)^{1/4}}$. \cite{lake} calculated the CM energy at the Cauchy horizon of a static Reissner-Nordstr{\o}m  black hole which is limited. \cite{grib} investigated the CM energy using multiple scattering process. The collision in the ISCO particles was investigated by the \cite{harada} for Kerr black hole.  \cite{liu} studied the BSW effect for Kerr-Taub-Nut space-times and showed that the CM energy
depends upon  both the Kerr parameter ($a$) and  the Nut parameter ($n$)  of the space-time.
\cite{mc} studied that the black holes are neither particle accelerators nor dark matter
probes. \cite{gala} also demonstrated that the center-of-mass  energy in the context of the near horizon geometry of the extremal Kerr black holes and proved that the center-of-mass energy is finite for any value of the particle parameters.
Several authors [\cite{wei,zaslav,patil}] have studied the BSW effect for different types  black holes and get the unlimited center-of-mass energy.

In this article, we wish to study the BSW effect of the Gibbons-Maeda-Garfinkle-Horowitz-Strominger (hereafter GMGHS) black hole and our aim is to observe what happens this effect precisely  in the extremal limit i.e. at $Q^{2}=2M^{2}e^{2\phi_{0}}$.

The dilatonic charged black hole are represented by the ``string metric" which is a solution of the effective action of the low energy limit of heterotic string theory can be written in the form:  \cite{gm,ghs}
\begin{eqnarray}
ds^2 &=& -\left(1-\frac{2M}{r}\right)dt^{2}+\left(1-\frac{2M}{r}\right)^{-1}dr^{2}
\nonumber \\[4mm] &&
+r\left(r-\frac{Q^{2}}{M}e^{-2\phi_{0}}\right)
[d\theta^{2}+\sin^{2}\theta d\phi^{2}] ~.\label{rn}
\end{eqnarray}
and
\begin{eqnarray}
e^{-2\phi} &=& e^{-2\phi_{0}}\left(1-\frac{Q^{2}}{Mr}e^{-2\phi_{0}}\right)\\
F &=& Q\sin\theta d\theta\wedge d\phi
\end{eqnarray}
where $\phi_{0}$ is the asymptotic value of the dilation field, $M$ represents the mass of the black hole, $Q$ denote its magnetic charge and $\phi$ is the scalar field.

Note that this metric differs from the Reissner-Nordstr{\o}m (RN) solutions of the Einstein-Maxwell theory is that it does not have any inner horizon. It also may be noted that this metric is exactly  similar to the Schwarzschild metric. But there is a difference in a sense that the area of the two sphere ($S^{2}$) of constant $r$ and $t$ now strictly depends upon the value of $Q$. This area ${\cal A}=4\pi r(r-b)$ (Here $r=2M, ~b=\frac{Q^{2}}{M}e^{-2\phi_{0}}$) goes to zero at the extremal limit i.e. at $b=2M$. The other interesting characteristics of this space-time is that there is a curvature singularity occurs at $r=b$. But from the string theoretical perspectives this singular nature manifested  when $b=2M$ which is not important because the string does not couple to the metric $g_{ab}$ rather to $e^{2\phi}g_{ab}$.

\section{ISCOs in GMGHS Black Hole}

In this section we will compute the the properties of the circular geodesics for GMGHS black hole in the equatorial plane and also compute the ISCOs for this black hole. To compute the geodesic motion of a test particle in this plane we set $\dot\theta=0$ and $\theta=constant=\pi/2$ and follow the reference \cite{pp2}.

Thus the  Lagrangian for this motion is given by
\begin{eqnarray}
2\cal L &=& -(1-\frac{2M}{r})\,(u^{t})^2+(1-\frac{2M}{r})^{-1}\,
(u^{r})^2 \nonumber \\[4mm] &&
+r(r-b)\,(u^{\phi})^2 ~.\label{lags}
\end{eqnarray}
The generalized momenta reads as
\begin{eqnarray}
p_{t} &=&-(1-\frac{2M}{r})\,u^{t}=-E =Const ~.\label{pts}\\
p_{\phi} &=& r(r-b)\,u^{\phi}=L=Const ~.\label{pps}\\
p_{r} &=& (1-\frac{2M}{r})^{-1}\, u^{r}  ~.\label{prs}
\end{eqnarray}
Here over dot represents differentiation with respect to proper time($\tau$). Since the Lagrangian of the test particle independent  of $`t'$ and `$\phi$', so $p_{t}$ and $p_{\phi}$ are conserved quantities.
Solving (\ref{pts}) and (\ref{pps}) for $u^{t}$ and $u^{\phi}$ we find
\begin{eqnarray}
u^{t}=\frac{E}{(1-2M/r)} ~.\label{tdot} \\
u^{\phi}=\frac{L}{r(r-b)}~.\label{phid}
\end{eqnarray}
where $E$ and $L$ are the energy and angular momentum per unit mass of the test particle.

Therefore the required Hamiltonian reads as
\begin{eqnarray}
\cal H &=& p_{t}\,u^{t}+p_{\phi}\,u^{\phi}+p_{r}\,u^{r}-\cal L ~.\label{hams}
\end{eqnarray}

In terms of the metric the Hamiltonian  may be written as
\begin{eqnarray}
\cal H &=&-(1-\frac{2M}{r})\,(u^{t})^{2}+(1-\frac{2M}{r})^{-1}(u^{r})^2
\nonumber \\[4mm] &&
+r(r-b)(u^\phi)^2-\cal L ~.\label{hams1}
\end{eqnarray}
Since the Hamiltonian is independent of $`t'$, therefore we can write it as
\begin{eqnarray}
2\cal H &=& -(1-\frac{2M}{r})\,(u^{t})^{2}+(1-\frac{2M}{r})^{-1}(u^{r})^2
\nonumber \\[4mm] &&
+r(r-b)(u^\phi)^2 ~.\label{ham1}\\
        &=&-E\,u^{t}+L\,u^{\phi}+\frac{1}{(1-2M/r)}\,(u^r)^2=\epsilon=const ~.\label{hs1}
\end{eqnarray}
Here $\epsilon=-1$ for time-like geodesics, $\epsilon=0$ for light-like geodesics and $\epsilon=+1$ for space-like geodesics.
Substituting the equations, (\ref{tdot}) and (\ref{phid}) in (\ref{hs1}), we obtain the radial equation for any spherically space-time is
\begin{eqnarray}
(u^{r})^{2}=E^{2}-{\cal V}_{eff}=E^{2}-\left(\frac{L^{2}}{r(r-b)}-\epsilon \right)(1-\frac{2M}{r})~.\label{radial}
\end{eqnarray}
where the standard effective potential for GMGHS space-time denoted as
\begin{eqnarray}
{\cal V}_{eff}=\left(\frac{L^{2}}{r(r-b)}-\epsilon \right)(1-\frac{2M}{r}) ~.\label{vrn}
\end{eqnarray}

To compute the circular geodesic motion of the test particle in the Einstein -Maxwell gravitational field , we must
have for circular geodesics of constant $r=r_{0}$  and from the equation (\ref{radial}) we have
\begin{eqnarray}
{\cal V}_{eff} &=& E^{2} ~.\label{v}
\end{eqnarray}
and
\begin{eqnarray}
\frac{d{\cal V}_{eff}}{dr} &=& 0 ~.\label{dvdr}
\end{eqnarray}
Thus we obtain the energy and angular momentum per unit mass of
the test particle are given by
\begin{eqnarray}
E^{2}_{0} &=&  \frac{(2r_{0}-b)(r_{0}-2M)^{2}}{r_{0}(2r_{0}^{2}-(b+6M)r_{0}+4Mb)} ~.\label{engg}
\end{eqnarray}
and
\begin{eqnarray}
L^{2} _{0} &=& \frac{2Mr_{0}(r_{0}-b)^2}{2r_{0}^{2}-(b+6M)r_{0}+4Mb}~ .\label{angg}
\end{eqnarray}

Circular motion of the test particle to be exists when both the energy and angular momentum are real finite, therefore we must have $2r_{0}^{2}-(b+6M)r_{0}+4Mb>0$ and $r_{0}>b$.

Again the angular frequency measured by an asymptotic observers for time-like circular geodesics at $r=r_{0}$ same as the angular frequency of Schwarzschild black hole which
is given by
\begin{equation}
\Omega_{0}=\frac{u^\phi}{u^t}=\sqrt{\frac{M}{r_{0}^3}}
\end{equation}
In general relativity, circular orbits do not exists for all values of $r$, so the denominator of equations (\ref{engg},\ref{angg}) real only if $2r_{0}^{2}-(b+6M)r_{0}+4Mb\geq 0$. The limiting case of equality gives an circular orbit with indefinite energy per unit
mass, i.e. a circular photon orbit. This photon orbit is the innermost boundary of the circular orbit for massive particles. It occurs at the radius
\begin{equation}
 r_{c} = \frac{1}{4}(b+6M+\sqrt{b^2-20Mb+36M^2})
\end{equation}

For extremal GMGHS black hole, the photon orbit occurs at $r_{c}=2M$. Marginally bound circular orbit(MBCO)  can be obtained by setting
$E=1$, then the radius of marginally bound orbit located at $r_{mb}=2M\pm\sqrt{2M(2M-b)}$. In the limit $b\rightarrow 0$, we get
$r_{mb}=4M$ which is the radius of marginally bound circular orbit of Schwarzschild black hole. At the extreme limit $b=2M$, marginally
bound circular orbit occurs at the radius $r_{mb}=2M$ for extremal GMGHS black hole.
From astrophysical viewpoint the most important class of orbits are  the innermost stable circular orbit(ISCO), which occurs at the
point of inflection of the effective potential ${\cal V}_{eff}$. Thus at the point of inflection
\begin{eqnarray}
\frac{d^2{\cal V}_{eff}}{dr^2}=0
\end{eqnarray}
with the auxiliary equation $\frac{d{\cal V}_{eff}}{dr}=0$. Then the ISCO equation for GMGHS black hole is given by
\begin{eqnarray}
 r_{0}^3-6Mr_{0}^2+6Mbr_{0}-2Mb^2=0
\end{eqnarray}
The real positive root of the equation gives the radius of ISCO at $r_{0}=r_{ISCO}$ which is given by
\begin{eqnarray}
 \frac{r_{ISCO}}{M} &=& 2+Z+\frac{2\left(2-\frac{b}{M}\right)}{Z}\\
\end{eqnarray}
where
\begin{eqnarray}
Z=\left[8-6(\frac{b}{M})+(\frac{b}{M})^{2}+ \sqrt{(\frac{b}{M})^{4}-4(\frac{b}{M})^{3}+4(\frac{b}{M})^{2}}\right]^{\frac{1}{3}}
\end{eqnarray}
Now we turn to the derivation of CM energy for the string black hole.

\section{CM energy of the collision near the horizon of the string black hole:}

In this section, we shall compute the energy in the center-of-mass frame for the collision
of two neutral particles coming from infinity with $\frac{E_{1}}{m_{0}}=\frac{E_{2}}{m_{0}}=1$ and approaching the black hole with different angular momenta $L_{1}$ and $L_{2}$. The center of mass energy is derived by using the formula\cite{bsw} which is valid in both flat and curved space-time reads
\begin{eqnarray}
\left(\frac{E_{cm}}{\sqrt{2}m_{0}}\right)^{2} &=&  1-g_{\mu\nu}u^{\mu}_{1}u^{\nu}_{2}~.\label{cm}
\end{eqnarray}
where $u^{\mu}_{1}$ and $u^{\nu}_{2}$ are the 4-velocities of the two particles, which can be determine from the following equation(\ref{utur}).

For this we need to derive the four velocity of the colliding particles. We assume throughout this work the geodesic motion of the colliding particles are confined to the equatorial plane. Since the space-time has a time-like isometry generated by the
time-like Killing vector field $\xi$ whose projection along the four velocity $\bf u$ of geodesics $\xi.\bf u=- E$, is conserved along such geodesics(where $\xi\equiv \partial_{t}$). Similarly there is also the `angular momentum' $L=\chi.\bf u$ is conserved due to the rotational symmetry(where $\chi\equiv \partial_{\phi})$. Thus for the massive particles the components of the four velocity are
\begin{eqnarray}
  u^{t} &=& \frac{{E}}{f(r)}  \\
  u^{r} &=& \pm \sqrt{E^{2}-f(r)\left(1+\frac{L^{2}}{r(r-b)}\right)} \label{eff}\\
  u^{\theta} &=& 0 \\
  u^{\phi} &=& \frac{L}{r(r-b)} ~.\label{utur}
\end{eqnarray}
where $f(r)=1-\frac{2M}{r}$.
\begin{eqnarray}
u^{\mu}_{1}= \left(\frac{E_{1}}{f(r)},~ -X_{1},~ 0,~\frac{L_{1}}{r^{2}}\right) ~.\label{u1}\\
u^{\mu}_{2}= \left(\frac{E_{2}}{f(r)},~ -X_{2},~ 0,~\frac{L_{2}}{r^{2}}\right) ~.\label{u2}
\end{eqnarray}
Therefore using(\ref{cm}), we obtain the center of mass energy for this collision:
\begin{eqnarray}
\left(\frac{E_{cm}}{\sqrt{2}m_{0}}\right)^{2} &=&  1 +\frac{E_{1}E_{2}}{f(r)}
-\frac{X_{1}X_{2}}{f(r)}-\frac{L_{1}L_{2}}{r(r-b)} ~.\label{cm1} \\
\mbox{where}\\\nonumber
X_{1} &=& \sqrt{E_{1}^{2}-f(r)\left(1+\frac{L_{1}^{2}}{r(r-b)}\right)} \\
X_{2} &=& \sqrt{E_{2}^{2}-f(r)\left(1+\frac{L_{2}^{2}}{r(r-b)}\right)}
\end{eqnarray}
As we have assumed $E_{1}=E_{2}=1$ previously and substituting $f(r)=1-\frac{2M}{r}$ we obtain finally the center of mass energy near the horizon:
\begin{eqnarray}
E_{cm} = \sqrt{2}m_{0}\sqrt{\frac{8M(2M-b)+(L_{1}-L_{2})^{2}}{2M(2M-b)}} ~.\label{cm2}
\end{eqnarray}
As we know\cite{bsw}, the maximum center-of-mass energy strictly depends upon the value of critical angular momentum such that the particles can reach the event horizon with maximum tangential velocity. Now we compute the critical angular momentum for this
black holes in the following way.

In fact, the critical angular momentum and the critical radius can be determined from the effective potential. From the equation (\ref{radial}), at the critical point the effective potential satisfied the conditions as we defined previously by the equations
(\ref{v}) and (\ref{dvdr}). Therefore using these conditions we get the critical angular momentum for geodesics falling in is
\begin{eqnarray}
-\left(2M+\sqrt{4M^2-2bM}\right)\le L \le \left(2M+\sqrt{4M^2-2bM}\right). \nonumber
\end{eqnarray}

In the limit $b\rightarrow 0$, this yields the critical angular momentum for Schwarzschild black hole and the critical values are $\pm 4M$.

In the limit $b\rightarrow 0$, the above expression reduce to
\begin{eqnarray}
E_{cm} &=& \sqrt{2}m_{0}\sqrt{\frac{16M^{2}+(L_{1}-L_{2})^{2}}{4M^{2}}}~.\label{cmsch}
\end{eqnarray}
which is the CM energy for Schwarzschild black hole.

For non-extremal GMGHS spacetimes the CM energy reads as
\begin{eqnarray}
E_{cm} = \sqrt{2}m_{0}\sqrt{\frac{8M\left(2M-\frac{Q^{2}}{M}e^{-2\phi_{0}}\right)
+(L_{1}-L_{2})^{2}}{2M\left(2M-\frac{Q^{2}}{M}e^{-2\phi_{0}}\right)}}
\end{eqnarray}
which shows that the CM energy is finite and depends upon the asymptotic
value of the dilation field ($\phi_{0}$).

Whenever we taking the extremal limit $b=2M$ we get the CM energy near the
event horizon $r=2M$
\begin{eqnarray}
E_{cm} &=&  \sqrt{2}m_{0}\sqrt{\frac{8M(2M-b)+(L_{1}-L_{2})^{2}}{2M(2M-b)}} \\
E_{cm} &\longmapsto &  \infty \nonumber\\
\end{eqnarray}
which implies that the center-of-mass energy of collision  for extremal dilation black hole blows up as we approaches the extremal limit. Thus we get the unlimited C.M. energies. In fact it is independent of the critical values of the angular momentum. This is one of the main results of the paper.

Another important point should be noted here that for extremal GMGHS space-time \cite{pp2}
the ISCO, circular photon orbit and marginally bound circular orbit coincide with the event
horizon i.e. $r_{ISCO}=r_{ph}=r_{mb}=r_{hor}=2M$.
If we choose the different collision point say ISCO or photon orbit or marginally
bound orbit, then the center-of-mass energy  will be unlimited for each collision
point. This is an another interesting features of this space-time.

\section{CM Energy of the collision Near the Horizon of the RN Black Hole}

In this section we shall compute the  particle acceleration and collisions near the outer horizon in the Reissner Nordstr{\o}m space-time. Here we shall choose the different collision points, first we choose the collision point at near the event horizon which
is a surface of infinite red-shift, the second collision point will be at Cauchy horizon and finally we choose the collision point at ISCO. Now we compare the energy obtained for the different collision points.

\subsection{Review of Geodesic Motion of RN Spacetime:}

The well known metric of RN space-time  which is a static, asymptotically flat and spherically symmetric solutions of Einstein- Maxwell  equation is given by [\cite{sch}]

\begin{eqnarray}
ds^2 &=& -\left(1-\frac{2M}{r}+\frac{Q^{2}}{r^{2}}\right)dt^{2}+\left(1-\frac{2M}{r}+\frac{Q^{2}}{r^{2}}
\right)^{-1}dr^{2} \nonumber \\[4mm] &&
+r^{2}\left(d\theta^{2}+\sin^{2}\theta d\phi^{2}\right) ~.\label{rn}
\end{eqnarray}

The black hole has event horizon  which  is located at $r_{+}=M+\sqrt{M^{2}-Q^{2}}$ and Cauchy horizon which is located at $r_{-}=M-\sqrt{M^{2}-Q^{2}}$.  Let us consider $x^{\mu}(\tau)$ represents the trajectory of the moving particles.
$\tau$ is the proper time of the moving particles. We restrict ourselves the geodesic  motion of the particles confined on the equatorial plane i.e. $u^{\theta}=0$ or $\theta=\pi/2$. Thus the equatorial time-like geodesics for RN spacetimes are

\begin{eqnarray}
  u^{t} &=& \frac{{E}}{g(r)}  \\
  u^{r} &=& \pm \sqrt{E^{2}-g(r)\left(1+\frac{L^{2}}{r^2}\right)} \\
  u^{\theta} &=& 0 \\
  u^{\phi} &=& \frac{L}{r^2} ~.\label{utur1}
\end{eqnarray}

where $g(r)=1-\frac{2M}{r}+\frac{Q^2}{r^2}$, $E$ and $L$ are represents the energy and angular momentum per unit rest mass of the particle. The radial equation for RN space-time can be rewritten as in terms of standard effective potential
\begin{eqnarray}
(u^{r})^{2}=E^{2}-{\cal V}_{eff}=E^{2}-\left(1+\frac{L^{2}}{r^2} \right)\left(1-\frac{2M}{r}+\frac{Q^2}{r^2}\right)~.\label{rRN}
\end{eqnarray}
where the standard effective potential for RN space-time is given by
\begin{eqnarray}
{\cal V}_{eff}=\left(1+\frac{L^{2}}{r^2}\right)\left(1-\frac{2M}{r}+\frac{Q^2}{r^2}\right) ~.\label{vrn1}
\end{eqnarray}

Let us now compute the CM energy for the two colliding particles of the same rest mass $m_{0}$ in arbitrary Einstein-Maxwell gravitational field . It can be computed by using the formula as given in\cite{bsw}:

\begin{eqnarray}
\left(\frac{E_{cm}}{\sqrt{2}m_{0}}\right)^{2} &=&  1-g_{\mu\nu}u^{\mu}_{1}u^{\nu}_{2}~.\label{cm3}
\end{eqnarray}
where $u^{\mu}_{1}$ and $u^{\nu}_{2}$ are the 4-velocities of the two particles, which can be find from the following equation(\ref{utur1}).

\begin{eqnarray}
u^{\mu}_{1} = \left( \frac{E_{1}}{g(r)},~ -Y_{1},~ 0,~\frac{L_{1}}{r^{2}}\right)
~.\label{u3}\\
u^{\mu}_{2} = \left( \frac{E_{2}}{g(r)},~ -Y_{2},~ 0,~\frac{L_{2}}{r^{2}}\right)
~.\label{u4}
\end{eqnarray}

Therefore using(\ref{cm3}) one can obtain the center-of-mass
energy for this collision:
\begin{eqnarray}
\left(\frac{E_{cm}}{\sqrt{2}m_{0}}\right)^{2} &=&  1 +\frac{E_{1}E_{2}}{g(r)}
-\frac{Y_{1}Y_{2}}{g(r)}-\frac{L_{1}L_{2}}{r^2}~.\label{cm4}\\
\mbox{where}\\ \nonumber
Y_{1} &=& \sqrt{E_{1}^{2}-g(r)\left(1+\frac{L_{1}^{2}}{r^2}\right)}\\
Y_{2} &=& \sqrt{E_{2}^{2}-g(r)\left(1+\frac{L_{2}^{2}}{r^2}\right)}
\end{eqnarray}

As we assume $E_{1}=E_{2}=1$ previously and substituting $g(r)=1-\frac{2M}{r}+\frac{Q^2}{r^2}$ one could  obtain finally the center-of-mass energy near the  event horizon($r_{+}$) for
non-extremal RN space-time
\begin{eqnarray}
E_{cm}\mid_{r\rightarrow r_{+}} &=& \sqrt{2}m_{0}\sqrt{\frac{4r_{+}^2+(L_{1}-L_{2})^{2}}{2r_{+}^2}} ~.\label{cm5}
\end{eqnarray}

Near the Cauchy horizon the CM energy for RN space-time is given by
\begin{eqnarray}
E_{cm} \mid_{r\rightarrow r_{-}} &=& \sqrt{2}m_{0}\sqrt{\frac{4r_{-}^2+(L_{1}-L_{2})^{2}}{2r_{-}^2}} ~.\label{cm6}
\end{eqnarray}

In the near extremal limit $Q^2=M^2(1-\epsilon^2)$ for  RN black hole, one obtains the CM energy
\begin{eqnarray}
E_{cm} \mid_{r\rightarrow {1+\epsilon}} &=& \sqrt{2}m_{0}\sqrt{\frac{4(1+\epsilon)^2+(L_{1}-L_{2})^{2}}{2(1+\epsilon)^2}}
~.\label{cm7}
\end{eqnarray}

For the extremal RN space-time the value corresponds to CM energy near the horizon is
\begin{eqnarray}
E_{cm}\mid_{r\rightarrow M} &=& \sqrt{2}m_{0}\sqrt{\frac{4M^2+(L_{1}-L_{2})^{2}}{2M^2}} ~.\label{cm8}
\end{eqnarray}
We know that for extremal   RN black hole the ISCO is at $r=4M$ [\cite{ms,pp1}].
Thus If we consider the collision point to be ISCO then the corresponding value of CM energy 
on the ISCO for extremal RN space-time is given by
\begin{eqnarray}
E_{cm}\mid_{r\rightarrow 4M} = \nonumber \\
\sqrt{2}m_{0}\sqrt{\frac{9(26 M^2-L_{1}L_{2})-\sqrt{112M^2-9L_{1}^2}\sqrt{112M^2-9L_{2}^2}}
{144M^2}} ~.\label{cm9}
\end{eqnarray}

\section{CM Energy of the collision on the ISCO of the Schwarzschild Black Hole}

Here we shall extend our analysis for Schwarzschild black hole and compute the CM energy of the colliding particles on the ISCO.
Since the event horizon of the black hole is a surface of infinite blue-shift and here we shall show how much it different in the value
of CM energy if we choose the  collision point to be ISCO. This is the main aim of this section. Again we have already described the
space-time has symmetry namely time translation symmetry and rotational symmetry. Therefore the energy and angular momentum are
conserved quantities along the geodesics. They are denoted by $E$ and $L$. We also confined ourselves on the equatorial plane, so on that plane the equatorial time-like geodesics are described by
\begin{eqnarray}
  u^{t} &=& \frac{{E}}{1-\frac{2M}{r}}  \\
  u^{r} &=& \pm \sqrt{E^{2}-\left(1-\frac{2M}{r}\right)\left(1+\frac{L^{2}}{r^2}\right)} \\
  u^{\theta} &=& 0 \\
  u^{\phi} &=& \frac{L}{r^2} ~.\label{utur4}
\end{eqnarray}
and the components of four velocities are

\begin{eqnarray}
u^{\mu}_{1} = \left( \frac{E_{1}}{1-\frac{2M}{r}},~ -Z_{1},~ 0,~\frac{L_{1}}{r^{2}}\right)
~.\label{u3}\\
u^{\mu}_{2} = \left( \frac{E_{2}}{1-\frac{2M}{r}},~ -Z_{2},~ 0,~\frac{L_{2}}{r^{2}}\right)
~.\label{u5}
\end{eqnarray}
Since the Schwarzschild space-time is the special case of RN space-time, therefore one can compute the CM energy using the equation (\ref{cm4}) as

\begin{eqnarray}
\left(\frac{E_{cm}}{\sqrt{2}m_{0}}\right)^{2} &=&  1 +\frac{E_{1}E_{2}}{\left(1-\frac{2M}{r}\right)}-
\frac{Z_{1}Z_{2}}{\left(1-\frac{2M}{r}\right)} -\frac{L_{1}L_{2}}{r^2}~.\label{cm10}\\
\mbox{where}\\\nonumber
Z_{1} &=& \sqrt{E_{1}^{2}-\left(1-\frac{2M}{r}\right)\left(1+\frac{L_{1}^{2}}{r^2}\right)}\\
Z_{2} &=& \sqrt{E_{2}^{2}-\left(1-\frac{2M}{r}\right)\left(1+\frac{L_{2}^{2}}{r^2}\right)}
\end{eqnarray}

As we setting $E_{1}=E_{2}=1$ previously, thus one may obtain the CM energy on the ISCO for Schwarzschild black hole is given by

\begin{eqnarray}
E_{cm}\mid_{r\rightarrow 6M}=  \nonumber \\ \sqrt{2}m_{0}\sqrt{\frac{(90M^2-L_{1}L_{2})-\sqrt{18M^2-L_{1}^2}\sqrt{18M^2-L_{2}^2}}
{36M^2}} ~.\label{cm11}
\end{eqnarray}

Again we have already discussed in the previously, the values of critical angular momentum for Schwarzschild black hole are $\pm 4M$. Thus the maximum CM energy occurs for the opposite values of angular momentum i.e. $L_{1}=4M$ and $L_{2}=-4M$.
Therefore the maximum CM energy on the ISCO for Schwarzschild black hole  for these values of angular momentum is
\begin{eqnarray}
E_{cm}\mid_{r\rightarrow 6M} &=&\sqrt{\frac{52}{9}}m_{0} ~.\label{cm12}
\end{eqnarray}
whereas the maximum CM energy at the near horizon is $2\sqrt{5}m_{0}$.

\section{Discussion}

In this note, we have investigated the collision of two different particles of different angular momentum  with same rest mass and falling towards the spherically symmetric massless dilation black hole and computed the center-of-mass energy for this black hole. Our analysis suggests that for non-extremal GMGHS space-time  the CM energy is finite and depends upon the asymptotic value of the field strength.

For extremal GMGHS black holes the CM energy is unlimited  and independent of the extreme fine-tuning of the angular momentum of the colliding particles.  The another feature of this work is that for extremal GMGHS space-time three orbits namely ISCO, photon orbit and marginally bound orbit coalesce to the horizon \cite{pp2} thus we may vary the collision point but they gives the identical results (infinite energy). Thus in summary, for extremal GMGHS space-time it is shown that the center of mass energy of collision at $r \equiv r_{ISCO}=r_{ph}=r_{mb}=r_{hor}=2M$ is arbitrarily large.

\end{document}